\documentclass[a4paper]{spie}  
\usepackage[]{graphicx}

\title{Echo Tomography of Black Hole Accretion Flows \supit{a}}

\author{Keith Horne
\supit{b}
\skiplinehalf
University of St.Andrews, KY16~9SS, Scotland, UK \\
}

\authorinfo{
\supit{a} To appear in:
 {\it SPIE. Astronomical Telescopes and Instrumentation.
Future EUV-UV and Visible Space Astrophysics Missions and 
Instrumentation.
Waikoloa HI. 22-28~Aug~2002.}
Eds. J.C.Blades, O.H.Siegmund.
SPIE Proc 4856.
\\
\supit{b} Email: kdh1@st-and.ac.uk,
Web: http://star-www.st-and.ac.uk/$\sim$kdh1
}
 
\def\cc{\ifmmode {\rm cm}^{-3} \else cm$^{-3}$\fi}  
\def\kms{\ifmmode {\rm km\ s}^{-1} \else km s$^{-1}$\fi} 
\def\nH	{\ifmmode n \else $n$ \fi}
\def\NH	{\ifmmode N \else $N$ \fi}
\def\La {\ifmmode {\rm Ly}\alpha \else Ly$\alpha$ \fi}
\def\Ha {\ifmmode {\rm H}\alpha \else H$\alpha$ \fi}
\def\Hb {\ifmmode {\rm H}\beta \else H$\beta$ \fi}
\def\HeI {\ifmmode {\rm He}\,{\sc i} \else He\,{\sc i} \fi}
\def\HeII {\ifmmode {\rm He}\,{\sc ii} \else He\,{\sc ii} \fi}
\def\MgII {\ifmmode {\rm Mg}\,{\sc ii} \else Mg\,{\sc ii} \fi}
\def\NV {\ifmmode {\rm N}\,{\sc v} \else N\,{\sc v} \fi}
\def\SiIV {\ifmmode {\rm Si}\,{\sc iv} \else Si\,{\sc iv} \fi}
\def\CIV {\ifmmode {\rm C}\,{\sc iv} \else C\,{\sc iv} \fi}
\def\CIIIb {\ifmmode {\rm C}\,{\sc iii}\,] \else  C\,{\sc iii}\,] \fi}
\def\OIIIb {\ifmmode {\rm O}\,{\sc iii}\,] \else  O\,{\sc iii}\,] \fi}

\begin{document} 

  \maketitle 

\begin{abstract}

We discuss technologies for micro-arcsec
echo mapping of black hole accretion flows
in Active Galactic Nuclei (AGN).
Echo mapping employs time delays, Doppler shifts,
and photoionisation physics
to map the geometry, kinematics,
and physical conditions in the reprocessing region
close to a compact time-variable source of ionizing radiation.
Time delay maps are derived from detailed analysis of
variations in lightcurves at different wavelengths.
Echo mapping is a maturing technology at a stage of development
similar to that of radio interferometry just before the VLA.
The first important results are in,
confirming the basic assumptions of the method,
measuring the sizes of AGN emission line regions, 
delivering dozens of black hole masses,
and showing the promise of the technique.
Resolution limits with existing AGN monitoring datasets
are typically $\sim$5-10 light days. This should improve
down to 1-2 light days in the next-generation echo mapping
experiments, using facilities like 
{\em Kronos} and {\em Robonet} that are designed for and dedicated to 
sustained spectroscopic monitoring.
A light day is 0.4 micro-arcsec at a redshift of 0.1, thus
echo mapping probes regions $10^3$ times smaller than VLBI,
and $10^5$ times smaller than {\it HST}.

\end{abstract}


\section{Micro-Arcsec Echo Tomography}
\label{sec:introduction}

Angular resolution is a discovery frontier that drives
the development of astronomical technology.
{\it HST} delighted the public and transformed astronomy by
delivering stunning sub-arcsec images
at ultraviolet, optical, and more recently infrared wavelengths.
Large ground-based telescopes can now produce sub-arcsec
images at infrared wavelengths by using adaptive optics to
compensate for atmospheric turbulence.
Sub-arcsec radio maps are routinely constructed
from complex visibility measurements obtained with
multi-element radio interferometers ({\it VLA}, {\it Merlin}),
and milli-arcsec resolution is achieved
with the Very Long Baseline Array ({\it VLBA}).
The most ambitious technology on the angular resolution horizon
is the Micro-Arcsec X-ray Interferometry Mission,
{\it MAXIM}, which may one day image
the flow of material into nearby
black hole event horizons.

Today we are already exploring the
micro-arcsec structure of black hole accretion flows
through echo tomography experiments.
Tomography is an indirect imaging technique
that recovers an image from measurements
of projections of that image.
Tomography is analogous to interferometry,
except that the measurements are of projections rather than
fourier components of the image.
Echo mapping employs
time-resolved spectrophotometry to record
spectral variations in which detailed
information on spatial and kinematic structure
is coded as time delays and Doppler shifts.
This article discusses echo tomography technologies
in use and under development for micro-arcsec
mapping of black hole accretion flows.

\subsection{Black Hole Accretion Flows}

Black holes come in two varieties:
stellar-mass holes ($M \sim 10 M_\odot$) in X-ray binaries,
and supermassive black holes ($M \sim 10^{6-9}M_\odot$)
in the nuclei of galaxies.
Accretion disks convey angular momentum outward as
matter spirals inward to feed the central black hole.
Accretion theory and black hole scaling laws can be tested
by comparing phenomena in X-ray binaries and 
active galactic nuclei (AGN).
Although they are too small for direct or interferometric imaging, 
these accretion flows can be resolved by echo tomography.

Current models of AGN spectral energy distributions include
optical to X-ray emission from the accretion disk,
infrared emission from a dusty molecular torus
encircling and from some viewing angles obscuring the disk,
and relativistically beamed radio to X-ray emission from jets.
Emission lines arise from reprocessing of X-ray and EUV radiation
in two kinematically-distinct regions, 
the broad line region (BLR) with $V\sim10^4~\kms$,
and the narrow-line region (NLR) with $V<10^3~\kms$.
The NLR, resolved by {\it HST}, has a
bi-polar morphology interpreted as wide cones of ionising
radiation emerging from the unresolved nucleus.
The accretion disk and BLR are unresolved,
but these regions vary on timescales of days to months
in response to erratic changes in ionising radiation from
regions still closer to the black hole.
Light travel time within the system produces observable
time delays, providing indirect information on
the size and structure of the disk and BLR,
which make these regions accessible to echo tomography.

\subsection{Time Delay Paraboloids}

\begin{figure}[t]
\begin{center}
\begin{tabular}{c}
\includegraphics[height=10cm,angle=-90]{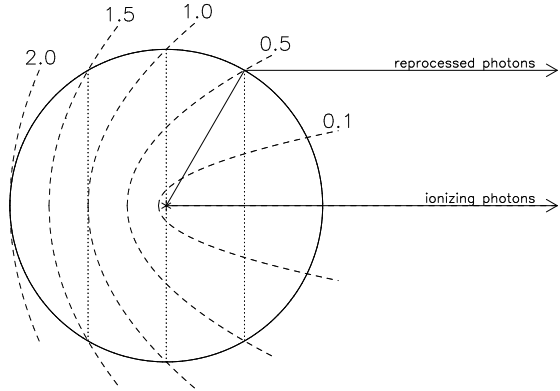} 
\end{tabular}
\end{center}
\caption{ 
Ionizing photons from a compact source are reprocessed
by gas clouds in a thin spherical shell.
A distant observer sees the reprocessed photons arrive
with a time delay ranging from 0 for reprocessing
at the near edge of the shell
to $2R/c$ for reprocessing at the far edge.
The iso-delay paraboloids slice the shell
into zones with areas proportional to the range of delays.
}
\label{fig:shell}
\end{figure}

Reverberation mapping 
\cite{Blandford_McKee_1982} relies on
a compact erratically variable source of ionizing radiation
embedded within the region that we wish to probe.
Heated and photoionised gas reprocesses the ionizing radiation
into ultraviolet, optical and infrared continuum and line emission.
Variations in the central source launch spherical waves of
heating and ionization that expand at the speed of light through
the surrounding gas.
Each change in ionization triggers a corresponding change
in the reprocessed emission.
In AGNs, the reprocessing time (hours) is small compared
with light travel time (days), so that the delay seen by a
distant observer is dominated by the light travel time.

The iso-delay surfaces are ellipsoids with one focus at the central source
and the other at the observer,
and are well approximated by paraboloids in the reprocessing 
region near the central source (Fig.~\ref{fig:shell}).
The time delay is
\begin{equation}
	\tau = \frac{R}{c} ( 1 + \cos{\theta} )\ ,
\end{equation}
for a reprocessing site at a radius $R$ from the nucleus
and azimuth $\theta$,
measured from 0 on the far side of the nucleus to $180^\circ$
on the line of sight between the nucleus and the observer.
The delay is 0 for gas on the line of sight between us and
the nucleus, and $2R/c$ for gas directly behind the nucleus
(Fig.~\ref{fig:shell}).
The time delay map, $\Psi(\tau)$,
is a 1-dimensional map of the emission line region,
effectively slicing up the region along
the nested set of iso-delay paraboloids.

\subsection{The Driving Lightcurve}

The ionizing radiation that drives reprocessing
includes EUV photons that are strongly absorbed by neutral
hydrogen in the interstellar medium,
and are therefore not directly observable.
Fortunately, we see nearly co-temporal variations in continuum
lightcurves at most ultraviolet and optical wavelengths,
suggesting that the continuum forms in a region
much smaller than the emission line regions.
The continuum light curve $f_c(t)$ thus serves as a useful
surrogate for the unobservable light curve of the ionizing radiation.

\subsection{ Linear and Linerized Reverberation Models}

In the simplest linear reverberation model,
reprocessing sites span a range of time delays,
and the line light curve $f_\ell(t)$ is a weighted
sum of time-delayed copies
of the continuum light curve $f_c(t)$,
\begin{equation}
	f_\ell(t) = \int_0^\infty 
		\Psi_\ell(\tau)\ f_c(t-\tau)\ d\tau .
\end{equation}
In this convolution integral, $\Psi_\ell(\tau)$ is 
the ``transfer function'', or ``convolution kernel'', or ``delay map''
of the emission line $\ell$.
This describes the strength of the reprocessed emission
that arises from the regions between pairs of iso-delay paraboloids.
Of course each emission line has its own time variations $f_\ell(t)$
and corresponding delay map $\Psi_\ell(\tau)$.

Echo mapping aims to recover $\Psi_\ell(\tau)$ from 
measurements of $f_\ell(t)$ and $f_c(t)$ made at specific times $t_i$.
To fit such observations,
the model above is too simple in several respects.
First, additional sources of light contribute
to the observed continuum and emission-line fluxes.
Examples are background starlight, and narrow emission lines.
Since these sources do not vary on the reverberation timescale,
they simply add constants to $f_\ell(t)$ and $f_c(t)$,
\begin{equation}
\label{eqn:linearized}
	f_\ell(t) = f_\ell^B  + \int_0^\infty 
		\Psi_\ell(\tau)\ 
		\left[ f_c(t-\tau) - f_c^B \right] 
		d\tau .
\end{equation}
This linearized model also useful as a tangent line approximation
to a non-linear line response.
The ``background'' fluxes, $f_\ell^B$ for the line
and $f_c^B$ for the continuum, are set somewhere near
the middle of the range of values spanned by the observations.
The delay map $\Psi_\ell(\tau)$ then gives the marginal response
of the line emission from gas at time delay $\tau$,
to changes in ionising radiation above or below
the chosen background level.

\subsection{Inversion Methods}

Three practical methods have been developed
for deriving $\Psi_\ell(\tau)$ from observations.
The Regularized Linear Inversion (RLI) method
\cite{Vio_Horne_Wamsteker_1994,Krolik_Done_1995}
and the
Subtractive Optimally-Localized Averages (SOLA) method
\cite{Pijpers_Wanders_1994},
use the linear reprocessing model 
to directly invert well-sampled equally-spaced
or interpolated lightcurves.
The Maximum Entropy Method (MEM),
is useful for the more sophisticated fitting problems,
allowing for non-uniform sampling and data quality
in the  lightcurves, and non-linear reprocessing models.
The echo mapping results presented in this paper are
obtained with the maximum entropy echo mapping code
\verb+MEMECHO+ 
\cite{Horne_Welsh_Peterson_1991,Horne_1994},
which makes use of the maximum entropy fitting code
\verb+MEMSYS+ 
\cite{Skilling_Bryan_1984}.

\section{First-Generation Echo Mapping Experiments}

\subsection{ Size and Radial Structure }

\begin{figure}[t]
\begin{center}
\begin{tabular}{c}
\includegraphics[height=10cm,angle=-90]{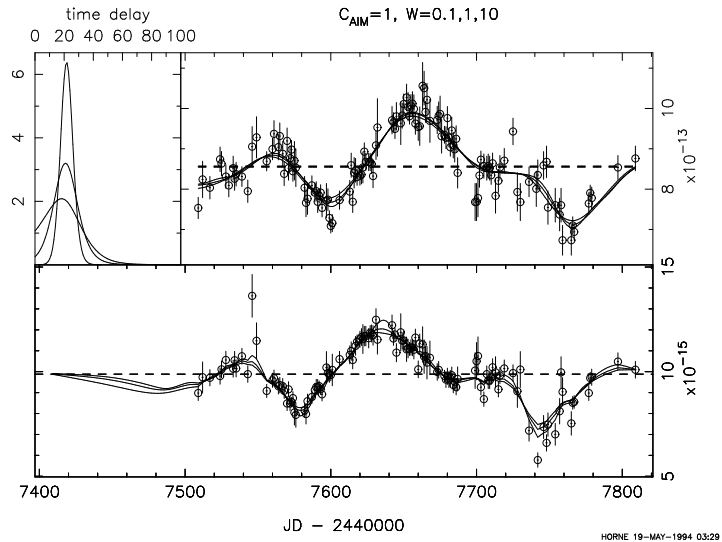} 
\end{tabular}
\end{center}
\caption{ 
Echo maps of \Hb\ emission in NGC~5548 
recovered by fitting to data from a 9-month AGN~Watch
monitoring campaign in 1989.
The optical continuum light curve (lower panel) is convolved with
the delay map (top left) to produce the \Hb\ emission-line
light curve (top right).
Horizontal dashed lines give the mean line and continuum fluxes.
Three fits are shown to indicate likely uncertainties.
Data from (Ref.~\cite{Horne_Welsh_Peterson_1991}).
}
\label{fig:hbmap}
\end{figure}

To illustrate the quality of the echo maps constructed
from current datasets,
Fig.~\ref{fig:hbmap} shows a \verb+MEMECHO+ fit
of the linearized echo model 
of Eqn.~(\ref{eqn:linearized})
to \Hb\ and optical continuum
lightcurves of NGC~5548 
\cite{Horne_Welsh_Peterson_1991}.
The lightcurves are extracted from optical spectra
arising from the 9-month {\it AGN~Watch} campaign in 1989.
Data from many observatories are inter-calibrated, and
the error bars are estimated from the internal consistency
of independent measurements made close together in time.
Subtracting the continuum background $f_c^B$,
convolving with the delay map $\Psi_\ell(\tau)$,
and adding the line background $f_\ell^B$,
gives the \Hb\ light curve.
The three fits shown, all with $\chi^2/N=1$,
give some impression of the uncertainty in the fit
arising from the noise level
of the data and gaps in time coverage.

The delay map of \Hb\ emission in NGC~5548 rises to a peak
at 20 days, and declines to low values by 40 days.
This measures the size of the \Hb\ emission-line
region, $\sim20$ light days.
Tests with simulated data using the same time sampling
and signal-to-noise ratios indicates that the resolution
achieved in this map is about 10 days.
Similar maps for a variety of ultraviolet emission lines
indicate that high ionisation lines have smaller delays than
low-ionisation lines
\cite{Krolik_etal_1991}.
This implies that reprocessing occurs over a
wide range of radii,
with higher ionisation closer to the nucleus.

\subsection{Velocity-Delay Maps }

\label{sec:kinematics}

To derive a Doppler-delay map from observations,
simply slice the observed spectra into wavelength bins,
and recover a delay map from the light curve at each
wavelength.
This is a simple extension of the echo model
used to fit continuum and emission-line lightcurves.
At each wavelength $\lambda$ and time $t$,
obtain the emission-line flux
\begin{equation}
	f_\ell(v,t) = f_\ell^B(v) +
	\int_0^\infty \left[ f_c(t-\tau) - f_c^B \right]\
	\Psi_\ell(v,\tau)\  d\tau
\end{equation}
by adding time-delayed copies of the continuum light curve
to a time-independent background spectrum $f_\ell^B(v)$.

\begin{figure}[t]
\begin{center}
\begin{tabular}{cc}
	\includegraphics[height=7cm,angle=-90]{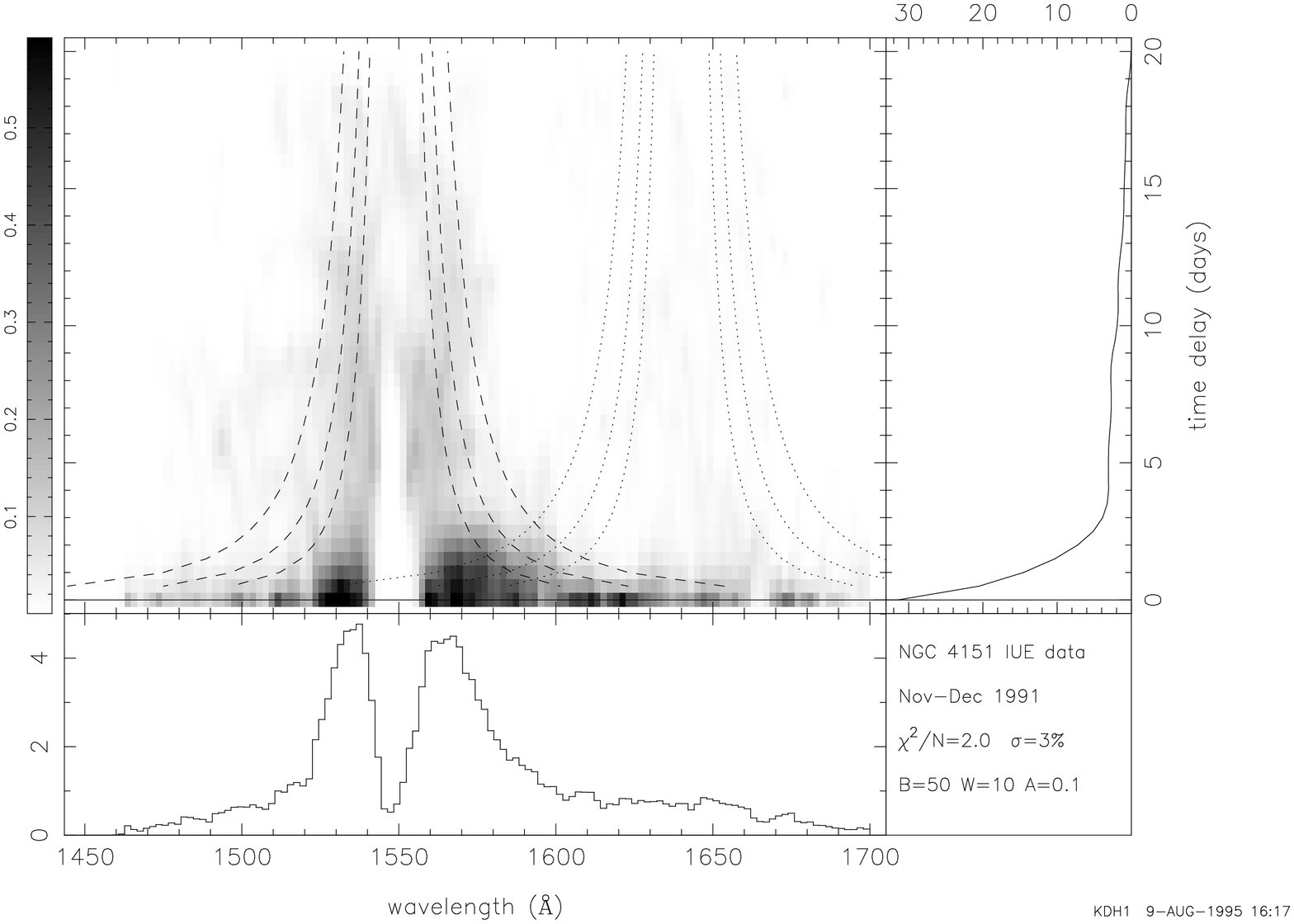} 
&
	\includegraphics[height=7cm,angle=-90]{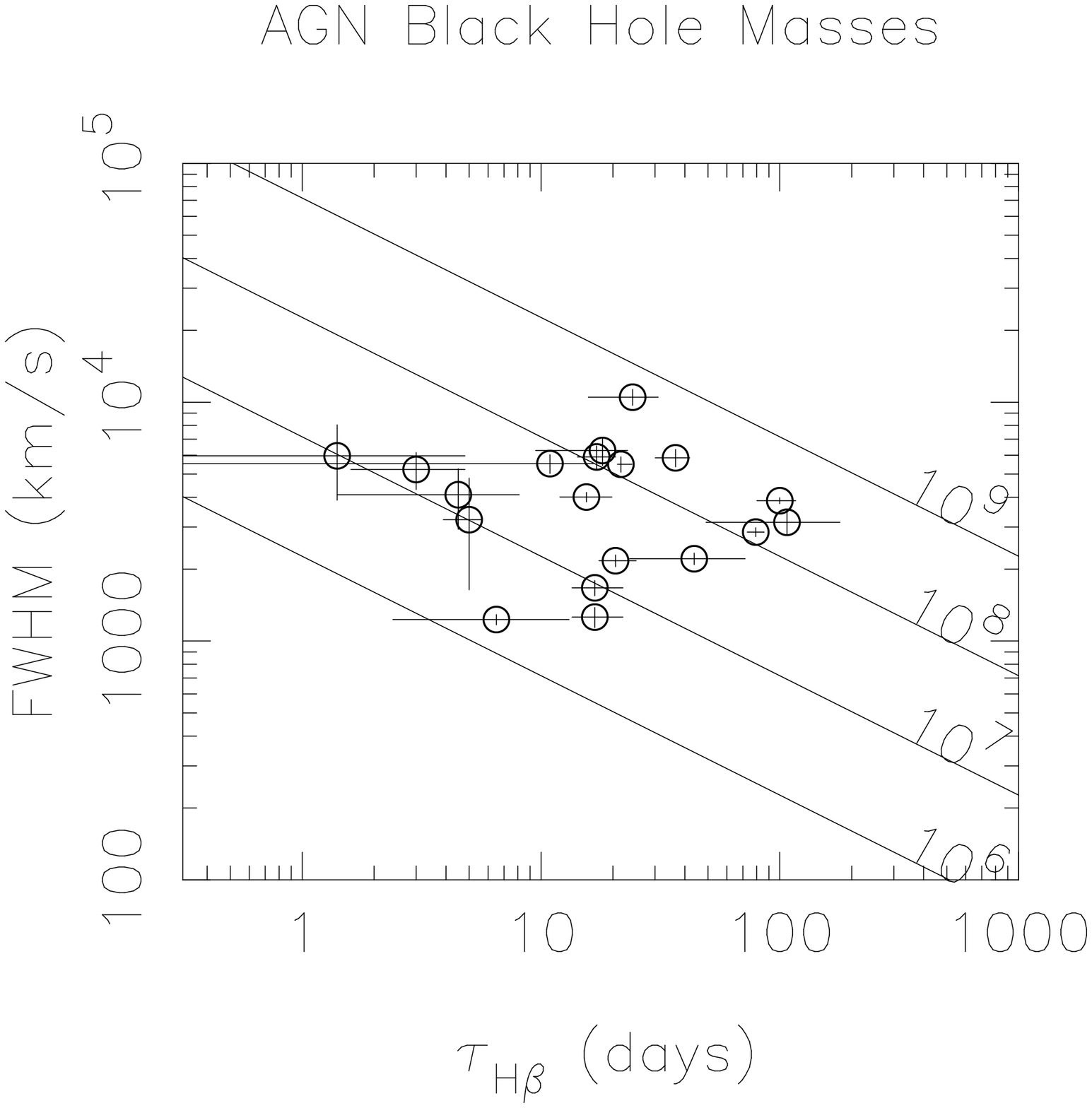} 
\end{tabular}
\end{center}
\caption{ Velocity-delay map of \CIV\ 1550 emission
(and blended \HeII\ 1640 emission) in NGC~4151
recovered by fitting to 44 IUE spectra.
Absorption wipes out the core of the  \CIV\ line.
Dashed curves give escape velocity envelopes
for masses
0.5, 1.0, and $2.0\times10^7$M$_\odot$.
An inverse correlation between velocity width and time delay
supports the hypothesis of virial motions, 
and yields virial mass estimates for 
dozens of supermassive black holes.
Black hole masses estimated from the \Hb\ 
emission line, based on cross-correlation
time delay and FWHM of the line in the
variable component of the light, 
are from 
(Ref.~\cite{Wandel_Peterson_Malkan_1999}).
}
\label{fig:4151}
\end{figure}

In Fig.~\ref{fig:4151}, the velocity-delay map for \CIV\ 1550
and \HeII\ 1640 emission was reconstructed from
44 {\it IUE} spectra of NGC~4151
\cite{Ulrich_Horne_1996}.
While spanning only 36 days, this campaign recorded
favorable continuum variations, including a bumpy exponential decline
followed by a rapid rise, which were sufficient to support echo
mapping on delays from 0 to 20 days.
Although a strong \CIV\ absorption feature
obliterates the delay structure at small velocities,
it is clear that a wide range of velocities is present
at small delays, and and smaller range at larger delays.
The dashed curves in Fig.~\ref{fig:4151} give escape
velocity envelopes
$v = \sqrt{2GM/c\tau}$ for masses 0.5, 1.0, and $2.0 \times
10^{7}$~M$_\odot$.
This clearly supports the hypothesis of virial motions,
and suggests a mass of order $10^{7}$~M$_\odot$.

\subsection{Black Hole Masses}

With the assumption of virial motions,
black hole mass can be estimated from
\begin{equation}
	M = f \frac{c\ \tau\ \sigma^2 }{G}
\ ,
\end{equation}
where $\tau$ and $\sigma$ are the time delay and velocity dispersion
of an emission line,
and $f$ is a form factor allowing for uncertain details of
the flow geometry, kinematics and orientation.
The virial hypothesis predicts that
lines formed at different radii in the same object should obey
$\sigma \propto \tau^{-1/2}$.
The internal consistency of mass estimates from different
lines in NGC~5548 
\cite{Krolik_etal_1991,Peterson_Wandel_1999},
and in a few other objects
\cite{Peterson_Wandel_2000},
supports the virial hypothesis.
An uncertainty of perhaps a factor of 3 remains
due to the factor $f$ representing ambiguity in the detailed
kinematics of the flow.
\cite{Krolik_2001}

Reverberation masses ranging from $10^{6}$ to $10^{9}$M$_\odot$
are now available for several dozen AGNs 
\cite{Wandel_Peterson_Malkan_1999,Kaspi_etal_2000}
based on cross-correlation time delays and rms widths
of \Hb emission (Fig.~\ref{fig:4151}).
At present there are no reverberation masses for 
high-redshift AGNs.
However, the low-redshift reverberation masses
serve to calibrate an empirical mass-radius-luminosity
relationship from which black hole masses may be estimated
for more distant systems 
\cite{Wandel_Peterson_Malkan_1999,Kaspi_etal_2000}.
This extends black hole mass estimates to all AGN with
a measured \Hb\ line width.
Such estimates are useful in the investigation of
the evolution of black hole masses and accretion rates.

\subsection{ Temperature Profiles of AGN Accretion Disks }

In theory, steady-state accretion disks have a
$T \propto (M \dot{M})^{1/4} R^{-3/4}$ structure,
and give rise to a characteristic spectrum
$f_\nu \propto  (M \dot{M})^{2/3} \lambda^{-1/3} D^{-2}$.
Here $\dot{M}$ is the accretion rate and $D$ is
the distance.
AGN spectra are generally much redder than this,
casting doubt on the validity of the disk model.
We can measure the $T(R)$ profiles of AGN accretion disks
by measuring $\lambda$-dependent continuum reverberations
arising from the disk surface.
With $T(R)$ decreasing outwards,
reverberations at smaller $R$ and shorter $\lambda$
will preceed those at larger $R$ and longer  $\lambda$.
Time delays increase as $\tau \approx R/c$, and
blackbody spectra peak at $T \approx h c / k \lambda$,
so that shorter wavelengths sense disk annuli at higher temperatures.
A disk surface with $T\propto R^{-b}$
will reverberates with a delay spectrum 
$\tau \propto \lambda^{-1/b}$.
The $T\propto  (M \dot{M})^{1/4} R^{-3/4}$ profile 
of a steady-state disk corresponds to 
$\tau \propto (M \dot{M})^{1/3} \lambda^{4/3}$.

AGN continuum lightcurves at different wavelengths 
exhibit a high degree of correlation, with time delays
generally shorter than a few days.
In the best-observed case
\cite{Collier_etal_1999}, NGC~7489, 
the  time delay does increase with wavelength,
consistent with $\tau \propto \lambda^{4/3}$.
Moreover, subtracting  bright and faint spectra
yields a difference spectrum close to
$f_\nu \propto \lambda^{-1/3}$.
These results suggest that
variability may be an effective way to isolate
the disk component of AGN spectra.
The  NGC~7489 results
yield estimates for $M \dot{M}$ and $D$,
leading to an encouragingly 
sensible estimate of the Hubble constant,
$H_0\sqrt{\cos{i}} \approx 42\pm9$\kms/Mpc.
If similar results are obtained for 
a larger sample of AGNs, spanning a range of redshifts,
the results may be used to to probe
cosmology beyond the redshift horizon of supernovae.

\subsection{Highlights of First-Generation Echo Mapping Experiments}

The first decade of echo mapping experiments,
a series of intensive monitoring campaigns undertaken
in particular by the {\it AGN~Watch} consortium
(\verb+http://www.astronomy.ohio-state.edu/~agnwatch/+),
has sharpened our knowledge of AGNs in numerous ways.
The basic photoionisation picture is well supported by
the correlated emission line and continuum variations,
with lines lagging behind the continuum,
in several dozen nearby AGNs.
Interpreting the time delay, $\tau$, as light travel time,
the inferred BLR sizes, $R \sim c\tau$, are
10-100 times smaller than expected
from earlier single-cloud photoionization models.
The smaller size implies higher densities, $\nH \sim 10^{11}\cc$.
Higher-ionization lines
have smaller delays and larger widths
\cite{Clavel_etal_1991},
suggesting a radially-stratified ionization structure
with virial kinematics\cite{Peterson_Wandel_1999}.
Larger time delays in more luminous sources imply
$R \propto L^{0.5-0.7}$,
compatible with an ionization-bounded BLR 
\cite{Kaspi_etal_2000}.
In several sources, an anti-correlation 
between velocity dispersions and time delays
is consistent with virial motions.
\cite{Krolik_etal_1991,Peterson_Wandel_1999,Peterson_Wandel_2000}
On this basis, ``reverberation masses'' 
are available for black holes in AGNs for which the
\Hb\ velocity dispersion $\sigma$ and 
time delays $\tau$ are measured.
This in turn calibrates an empirical $M$--$\sigma$--$L$
relationship from which black hole masses can be
estimated for all AGNs with measured \Hb\ line widths
\cite{Wandel_Peterson_Malkan_1999}.
We thus have a foundation for demographic studies of the
growth of supermassive black holes over cosmic time.

\section{ Next-Generation Echo Mapping Experiments}

In the next decade, striking improvements in echo mapping
capabilities are expected from technology
developments on two fronts.

First, improvements in the quality and time sampling of
spectrophotometric monitoring datasets should dramatically 
improve the resolution and fidelity of the maps.
The present 5-10 day resolution of echo maps is
limited by the noise levels (typically 3\%),
cadence, and duration of the lightcurves.
Just as the {\em VLA} dramatically improved the $U-V$ coverage 
and hence resolution and fidelity of radio maps,
so the echo maps will be greatly improved
by using facilities that are specifically designed
for and dedicated to sustained spectrophotometric monitoring.
\begin{itemize}
\item {\em RoboNet}
(\verb+http://star-www.st-and.ac.uk/~kdh1/jifpage.html+) is
a global network of 2-m robotic telescopes equipped with identical
multi-band CCD imagers and integral-field unit spectrographs.
\item Queue-scheduled spectrographs on large ground-based telescopes
are ideal for sustained daily monitoring,
and will enable echo mapping of fainter AGNs at larger redshifts.
\item {\em Kronos} (\verb+http://www.astronomy.ohio-state.edu/~kronos+)
is a small space telescope (NASA MidEx) 
equipped with simultaneous X-ray, UV, and optical spectrographs
in a 14~d orbit to enable nearly continuous coverage
for weeks to months and daily sampling for hundreds of days.
\end{itemize}
What may we expect to achieve in the
next-generation echo mapping experiments?
Simulation tests, some of which are presented below,
indicate that high fidelity echo maps
with a resolution exceeding 1 light day
will emerge from datasets with a time sampling $<1$~day,
a duration $\sim300$ days,
and accuracies of $\sim 1$\%.

Second, in tandem with better data,
better models are needed to interpret the improved datasets.
We currently extract lightcurves from observed spectra,
and fit those lightcurves with a simple 
linear or linearised reprocessing model
to construct delay maps for each line or wavelength
independently of the others.
This relatively model-independent approach
ignores a great deal of prior information.
Photoionisation models
predict highly non-linear and anisotropic responses that are
different for each emission line.
Given the evidence supporting the photoionisation hypothesis, 
we should now aim to improve the echo maps by building
in the additional constraints from photoionisation physics.
The next-generation reverberation models, currently under
development, will fit the entire reverberating spectrum,
avoiding the need to extract lightcurves and de-blend lines,
and build photoionisation physics into the fitting process.
This approach should yield 5-dimensional maps of the
geometry ($R$, $\theta$), kinematics ($V$), 
and physical conditions ($\nH$,$\NH$) in the photoionised gas.

We discuss and illustrate these developments below.

\subsection{ Better Datasets, Sharper Maps }

\begin{figure}[t]
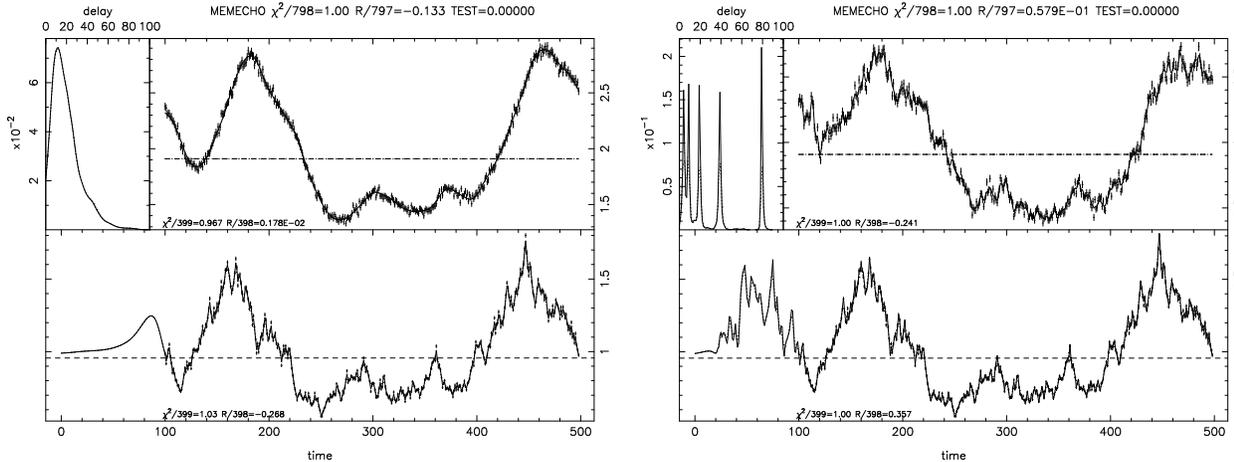

\begin{center}
\begin{tabular}{cc}
\includegraphics[height=8cm,angle=-90]{horne_fig4a.eps} 
&
\includegraphics[height=8cm,angle=-90]{horne_fig4b.eps} 
\end{tabular}
\end{center}
\caption{
Two simulations, each illustrating the recovery of a delay map (top left)
from data points sampling an erratically varying continuum light
curve (bottom) and the corresponding delay-smeared
emission-line light curve (top right).
The data points sample 400 epochs spaced by 1 day.
In simulation~1 (left) the true map is
$\Psi_\ell(\tau) \propto \tau e^{-\tau/10}$.
In simulation~2 (right)
the map has 5 sharp peaks
at delays 5, 10, 20, 40, and 80 days.
In both cases the reconstructed delay maps
closely resemble the true maps.
}
\label{fig:faketest}
\end{figure}

Fig.~\ref{fig:faketest}
illustrates the recovery of 1-dimensional delay maps based on 
\verb+MEMECHO+ fits to simulated lightcurves.
The time sampling and signal-to-noise ratios in these simulations
are designed to resemble datasets that will arise routinely
with next-generation spectrophotometric monitoring
facilities, such as {\em Kronos} and {\em RoboNet}.

In both simulations, examine and compare
the driving continuum lightcurve (bottom)
with the responding line lightcurve (top right).
The delayed maxima and minima indicate
that the delay map has a mean delay of about 20 days.
The fast variations evident in the continuum lightcurves 
appear to be washed out in the line lightcurves,
which are much smoother, indicating
that a range of delays is present.
By eye, this is about all that we can safely infer.
The line lightcurve in simulation~2 does exhibit more
fast structure than that in simulation~1,
but this structure is not cleanly correlated with
the fast continuum variations.
It is not obvious what this may imply about the delay structure.
Evidently our eyes and brains have a rather limited ability
to interpret reverberation datasets by inspection
-- much as is also the case for
$U-V$ visibility measurements that are employed in interferometry.

When \verb+MEMECHO+ fits the two lightcurves in detail, 
however, information
from all the small but significant changes 
recorded in the lightcurves
is assembled to construct the delay map (top left).
The two maps recovered by  \verb+MEMECHO+ are quite distinct.
In simulation~1 the delay map has a smooth distribution,
rising to a rounded peak and then declining smoothly.
In simulation~2 the delay map has 5 discrete peaks.
In both cases the recovered maps
closely resemble the true map.
With such high quality data,
the delay map is recovered with high fidelity.
There is some blurring, of course, because no finite
dataset can recover the map with infinite resolution.
The resolution achieved is about 1 light day.

\begin{figure}[t]
\begin{center}
\begin{tabular}{c}
	\includegraphics[height=10cm,angle=-90]{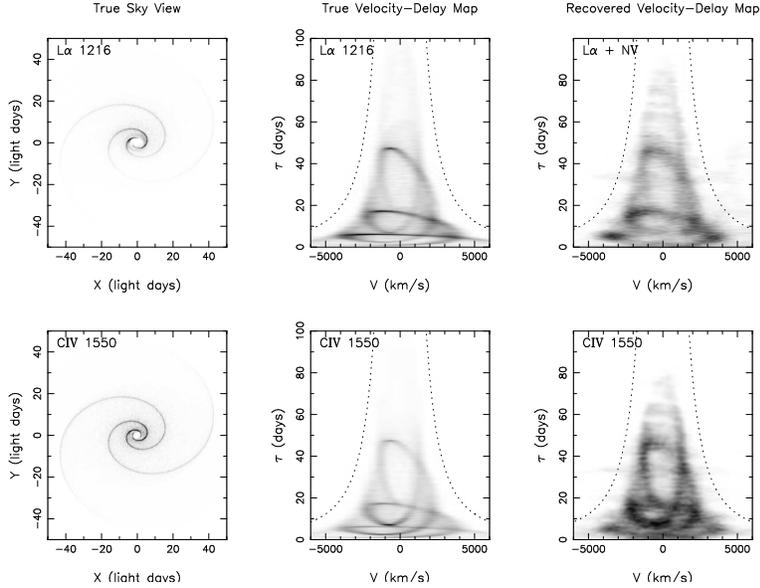} 
\end{tabular}
\end{center}
\caption{ High-fidelity velocity-delay maps of \La\ and \CIV\
emission reconstructed from a simulated {\em Kronos}
dataset spanning 200 days.
The emission is assumed to arise from a Keplerian accretion disk
with a two-armed trailing spiral density wave pattern (left).
The spiral waves
are clearly visible in the true velocity-delay map (middle)
and in the map recovered from the simulated {\em Kronos} spectra (right).
Dashed curves give the maximum time delay allowed at each
velocity for bound gas around a black hole of mass
$3\times10^7$M$_\odot$.
}
\label{fig:spiralmap}
\end{figure}

Fig.~\ref{fig:spiralmap} examines \verb+MEMECHO+ reconstructions of
2-dimensional velocity-delay maps from simulated {\em Kronos} datasets.
This simulation illustrates the extent to which
fine-scale structure in the geometry and kinematics of an AGN
emission line region can be recovered from a {\em Kronos} dataset.
The simulated dataset, designed to resemble a {\em Kronos} observation
of NGC~5548, includes realistic noise levels allowing for the effective 
area curves of the  {\em Kronos} spectrographs, and assuming that a 1 hour 
spectrum is taken every 4.8 hours for a total of 200 days.
We adopt the power-law model of Kaspi \& Netzer
\cite{Kaspi_Netzer_1999},
in which the density $\nH$ and column density $\NH$ are
power-law functions of radius,
and the photoionisation code \verb+ION+ is used
to evaluate the anisotropic and non-linear responses of the
emission lines at each level of ionising continuum flux.
For the geometry and kinematics we adopt a Keplerian disk 
with a 2-armed trailing spiral density wave.
The disk is comprised of
a large number of discrete clouds orbiting
a black hole of mass $3\times10^7$M$_\odot$.
The co-planar Keplerian orbits are elliptical, 
so that each cloud spends more time near apbothron than peribothron,
and are rotated by an angle proportional to the logarithm of 
their semi-major axes, thereby producing the spiral density wave.
From the synthesized spectra, we fit continua, extract line
lightcurves at each velocity, and use \verb+MEMECHO+
to construct velocity-delay maps for each line.

In Fig.~\ref{fig:spiralmap}, the spiral waves 
clearly visible on the sky view (left panel)
extend out to $\approx10^{-4}$ arcsec
(the redshift of NGC~5548 is $z=0.016$).
This structure is beyond the reach of
foreseeable developments in direct or 
interferometric imaging technology.
Yet the velocity-delay map (middle panel)
clearly shows this spiral wave structure,
and it is also clearly recognizable
in the reconstructed velocity-delay map (right panel)
recovered from the simulated {\em Kronos} dataset.

These results are typical of many other simulations
that we have run, for a wide variety of source structures,
indicating that major advances in the 
fidelity and information content of echo maps will arise
when the data quality is significantly improved.
The simulations indicate that we can sharpen our vision
by a factor of 10 compared to the 5-10 day resolution
that has been achieved to date.

\subsection{ 5-Dimensional Maps: $\Psi(R,\theta,\nH,\NH,v)$}
\label{sec:conditions}

In parallel with improvements in the reverberation datasets,
we must improve the models that we use to interpret the data.
A great deal of information on the geometry,
kinematics, and physical conditions of
black hole accretion flows is encoded in the detailed
reverberations of emission-line spectra.
We cannot extract this information by inspection of the data.
To tap this information, and 
reconstruct maps of black hole environments,
we must aim to fit the observed reverberations
in far greater detail than has previously been attempted.
The next-generation echo-mapping codes, building in
constraints from photoionisation physics,
are currently under development.
We sketch the new methods, and then illustrate
some preliminary results below.

Rather than extracting line and continuum lightcurves,
we now synthesize and fit $f_\nu(\lambda,t)$,
the complete spectrum at each time.
This avoids sticky problems of continuum fitting and
de-blending the overlapping wings of emission lines.
We model the spectrum as a sum of three components:
direct light from the nucleus,
reprocessed light from the surrounding gas clouds,
and background light:
\begin{equation}
	f_\nu(\lambda,t) =
	f^D_\nu(\lambda,t) + f^R_\nu(\lambda,t) + f^B_\nu(\lambda)
\ .
\end{equation}
The background light, $f^B_\nu(\lambda)$, allows for
contamination by non-variable sources, e.g. starlight
from the host galaxy.
The direct light from the nucleus is
\begin{equation}
	f^D_\nu(\lambda,t) =
	\frac { L(t) S_\nu(\lambda) } { 4 \pi D^2 }
\ ,
\end{equation}
where $D$ is the distance, $L(t)$ is the lightcurve,
and $S_\nu(\lambda)$ is the shape of the spectrum
emerging from the nucleus.
The emission-line spectrum of the reprocessed light is
\begin{equation}
f^R_\nu(\lambda,t)
	= \int_0^\infty 
	\frac{L(t-\tau)}{4\pi D^2}\
	\Psi_\nu( \lambda, \tau, L(t-\tau) )\
	d\tau
\ ,
\end{equation}
arising from a non-linear transfer function
given by
\begin{equation}
\begin{array}{rl}
\Psi_\nu( \lambda, \tau, L )
 & = \sum_\ell\ \int \int \int \int
2\pi R\ dR\ \sin{\theta}\ d\theta\ d\nH\
	d\NH\ dv\
\\ & \times 	
	\Psi(R,\theta,\nH,\NH,v)
\	\epsilon_\ell(F,\nH,\NH,\theta)
\\ & \times	g_\nu(\lambda-(1+v/c)\lambda_\ell)
\	\delta\left( 
		R - \left( \frac{ L } { 4 \pi F }	
	\right)^{1/2} \right)\
\	\delta\left(
		\tau - \frac{R}{c} \left( 1 + \cos{\theta} 
	\right)\right)\
\ .
\end{array}
\end{equation}
This at first sight frighteningly complicated expression
is simply a sum of contributions
over all the emission lines,
	$\ell$,
integrated over the geometric volume,
	$ 2\pi R\ dR\ \sin{\theta}\ d\theta$,
and the different types of gas clouds, 
	$ d\nH\ d\NH\ dv$.
The 5-dimensional map describing
the population of gas clouds
is the differential covering factor, 
$\Psi(R,\theta,\nH,\NH,v)$.
Each type of cloud is specified by 5 parameters,
the density $\nH$, 
column density $\NH$, 
distance from the nucleus $R$,
azimuth $\theta$, 
and Doppler shift $v$.
\footnote{The clouds also have a position angle $\phi$ around the line of 
sight,
and two perpendicular velocity components, but we omit them
here because the data are unchanged if we rotate
the cloud distribution around the line of sight.}
At the observers's time $t$,
a cloud at radius $R$ and azimuth $\theta$ is
exposed to an ionizing photon flux
\begin{equation}
	F(t,R,\theta) = \frac {L(t-\tau)} {4 \pi R^2 }
\ ,
\end{equation}
which allows for the light travel time delay
\begin{equation}
	\tau = \frac{R}{c} \left( 1 + \cos{\theta} \right)
\ .
\end{equation}
The irradiated cloud responds
by emitting line $\ell$ with an efficiency 
$\epsilon_\ell(F,\nH,\NH,\theta)$, which we evaluate
using a photoionisation code, e.g.\ \verb+CLOUDY+.
Each line has a different reprocessing efficiency,
depending on the incident ionising flux $F$,
the cloud density \nH\ and column \NH,
the viewing angle $\theta$
\footnote{The viewing angle $\theta$ is the same as the azimuth.},
and element abundances.
Finally, the Gaussian frequency distribution 
$g_\nu$ applies the Doppler shift,
where $\lambda_\ell$ is the rest wavelength of the line,
and the two Dirac $\delta$ distributions 
ensure that the correct
time delay $\tau$ and ionizing flux $F$
are used at each reprocessing site.

To fit the above model to observations of $f_\nu(\lambda,t)$,
we must adjust the distance $D$, the ionizing radiation light curve
$L(t)$,
the background spectrum $f^B_\nu(\lambda)$,
and the 5-D cloud map $\Psi(R,\theta,\nH,\NH,v)$.
Constraints available from reverberating emission-line spectra,
perhaps $\sim 10^4$ data points,
are insufficient to uniquely determine the 5-dimensional
cloud map, $\Psi$, which may have $\sim 10^{6}$ pixels.
Our computer code \verb+MEMBLR+ uses
the maximum entropy method to locate
the ``simplest'' cloud maps that fit the data.
Desktop computers are now fast enough to support
this type of detailed modelling and mapping of AGN emission regions.
Note that because every line provides a different weighted average,
a different projection, of the 5-dimensional cloud distribution,
we have a generalised form of tomography.


\subsection{ Recovery of a Hollow Shell Geometry }

\begin{figure}[t]

\begin{center}
\begin{tabular}{cc}
	\includegraphics[height=10cm]{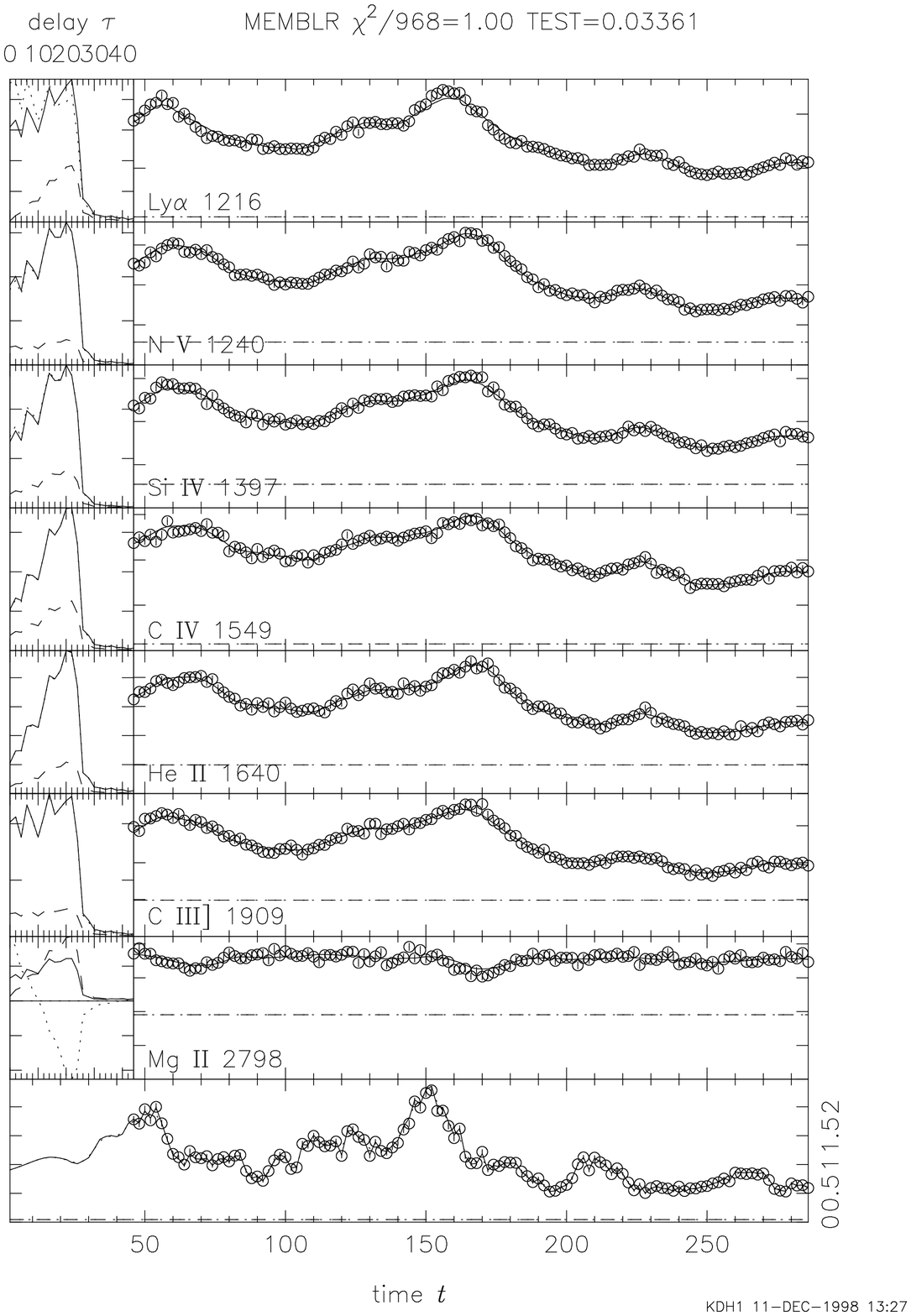} 
&	\includegraphics[height=10cm]{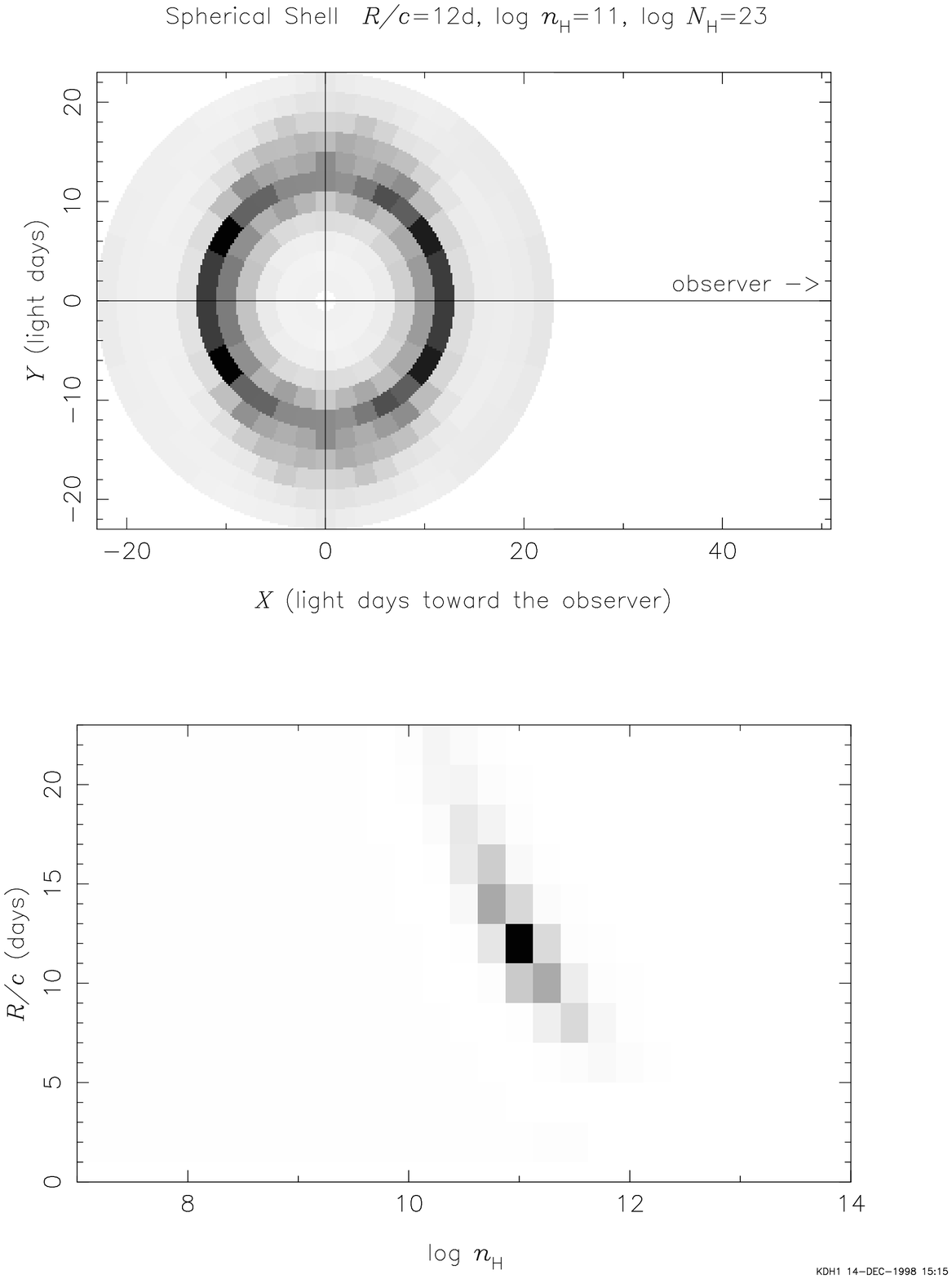} 
\end{tabular}
\end{center}
\caption{ 
Reconstructed map $\Psi(R,\theta,\nH)$
of a thin spherical shell (right) from a maximum entropy
fit to 7 ultraviolet emission-line lightcurves (left).
The light curve fits achieve $\chi^2/N=1$.
The shell radius $R/c = 12$d and density $\log{\nH}=11$
are correctly recovered.
For further details, see text.
}
\label{fig:shellfit}
\end{figure}

In Fig.~\ref{fig:shellfit} we present results of
a simulation test designed to investigate
how well the geometry and physical conditions 
may be recoverable from reverberations in ultraviolet
emission-lines.
In this simulation, we assume that clouds with density
$\nH=10^{11}$cm$^{-3}$ and column $\NH=10^{23}$cm$^{-2}$
are uniformly distributed over a thin spherical shell
of radius $R/c=12$d.
An erratically varying source at the centre of the hollow shell
drives reverberations in 7 ultraviolet emission lines.
We compute the emission-line lightcurves, allowing for
light travel time delays and 
using \verb+CLOUDY+ to account for each line's
anisotropic and non-linear reprocessing efficiency.
We sample the synthetic lightcurves at 121 epochs spaced by 2 
days, and add noise to simulate observational errors.
Finally, we reconstruct a 3-dimensional cloud map 
$\Psi(R,\theta,\nH)$
by fitting to the 7 synthetic lightcurves.

It is important to realise that
this \verb+MEMBLR+ fit does not assume a spherical shell geometry,
but rather it considers every possible cloud map,
$\Psi(R,\theta,\nH)$,
and tries to find the ``simplest'' map that fits the 
emission-line lightcurves.
The fit adjusts 4147 pixels in the cloud
map $\Psi(R,\theta,\nH)$,
143 points in the continuum light curve, $f_c(t)$, 
7 emission-line background fluxes, $f_\ell^B(\lambda)$, and
1 continuum flux, $f_c^B$.
This fit assumes the correct column density, $\NH$, and distance, $D$.
The fit to $N=968$ data points achieves $\chi^2/N=1$.
Entropy maximization ``steers'' each pixel in the map
toward its nearest neighbors, thus giving preference to ``smooth'' maps, 
and toward the pixel with the opposite sign of $\cos{\theta}$,
thus giving preference to maps with front-back symmetry.

On the left-hand side of Fig.~\ref{fig:shellfit}, we see that
the 8 lightcurves, 7 lines and 1 continuum, 
are well reproduced by the fit.
The highs and lows are a bit more extreme in the data --
a common characteristic of regularized fits.
To the left of each emission-line light curve, three
delay maps are shown corresponding to the maximum brightness
(solid curve), minimum brightness (dashed), 
and the difference (dotted).
All the lines exhibit an inward anisotropy except \CIIIb.
All the lines have positive linear responses except \MgII,
which has a positive response on the near side of the shell
and a negative response on the far side.

On the right-hand side of Fig.~\ref{fig:shellfit}, we see that the
fit to the 8 lightcurves recovers a hollow shell geometry
with the correct radius.
The density-radius projection of the map has a peak at the
correct radius and density.
The shell spreads in radius by a few light days,
with lower densities at larger radii, maintaining
a constant ionization parameter.

This simulation test suggests that reverberation effects
in the 7 ultraviolet emission-line lightcurves contain enough
information to construct useful maps of the geometry
and density structure of a photoionised emission-line zone.
The 3-dimensional cloud map $\Psi(R,\theta,\nH)$
clearly recovers the correct hollow shell geometry,
the correct radius, and the correct density.
The prospects are therefore good for detailed
mapping of the geometry and physical conditions in the BLRs of real AGNs.
While the current simulation fits only 7 emission-line lightcurves,
and demonstrates only 3-dimensional mapping,
it is straightforward, though more compute intensive,
to add  \NH\ and $v$ dimensions to the cloud map,
synthesize full spectra rather than just line fluxes,
and increase the number of lines to several hundred.
These extensions should increase the quality of the maps.

\section{Summary and Conclusion}
\label{sec:conclusion}

Echo tomography is being used to resolve AGN emission-line
regions on micro-arcsec scales.
In the first decade of echo mapping experiments,
datasets have been acquired with great effort through
international campaigns focusing on one object per year.
The main results are direct measurements of the sizes 
of broad emission-line regions,
several critical tests of photoionisation models
(radial ionization structure, luminosity-radius correlation),
and rough virial masses for dozens of supermassive black holes.

Our simulation tests illustrate some of the potential
future capabilities of echo mapping technology.
Future progress depends on facilities like
{\em RoboNet} and {\em Kronos}
that are designed for and dedicated to
spectrophotometric monitoring of AGNs (and other objects).
Delivery of continuous or at least daily records
of the evolving spectra over a few hundred days
will enable detailed and reliable mapping of
broad line regions in up to 5 dimensions,
including the geometry ($R,\theta$), 
kinematics ($v$),
and physical conditions ($n_H$, $N_H$).

Our simulations assume that the physical assumptions
and atomic data incorporated into the photoionisation models
are correct, and no doubt our knowledge of photoionisation
physics, embodied in codes like \verb+ION+ and \verb+CLOUDY+,
will continue to improve through increasingly
detailed confrontation with observations.
AGNs offer the opportunity to observe and test predictions for the
dynamic responses of photoionised gas.

Wavelength-dependent time delays in the AGN continuum should
also permit mapping the radial temperature profile $T(R)$
in the continuum production region, which is widely held to be
the surface of an accretion disk around the black hole.
This method could critically test the accretion disk hypothesis,
and may also provide a means of measuring AGN distances to
realize their potential as cosmological probes.

\acknowledgments     
 
KH is supported by a PPARC Senior Fellowship.


\bibliography{horne}   
\bibliographystyle{spiebib}   

\end{document}